\documentclass[12pt]{article}

\input epsf

\begin{document}

\title{Yang-Lee Zeros of the Q-state Potts Model on Recursive Lattices}
\author{R.G.~Ghulghazaryan$^1$, N.S.~Ananikian$^{1,2}$,  P.M.A.~Sloot$^3$ \\
{\small \textsl{$^1$ Department of Theoretical Physics, Yerevan Physics Institute,%
}} \\
{\small \textsl{Alikhanian Brothers 2, 375036 Yerevan, Armenia}} \\
{\small \textsl{$^2$ Dipartimento di Scienze Chimiche, Fisiche e Matematiche,}} \\
{\small \textsl{ Universita' degli Studi dell'Unsubria, via Vallegio, 11-22100 Como, Italy}} \\
{\small \textsl{$^3$ Section Computational Science, University of Amsterdam,}} \\
{\small \textsl{Kruislaan 403, 1098 SJ Amsterdam, The Netherlands}}}

\maketitle

\begin{abstract}

The Yang-Lee zeros of the Q-state Potts model on recursive lattices are studied 
for non-integer values of Q. Considering 1D lattice as a Bethe lattice with
coordination number equal to two, the location of Yang-Lee zeros of 1D
ferromagnetic and antiferromagnetic Potts models is completely analyzed 
in terms of neutral periodical points. 
Three different regimes for Yang-Lee zeros are found for $Q>1$ and $0<Q<1$.
An exact analytical formula for the equation of phase transition points is
derived for the 1D case. It is shown that Yang-Lee zeros of the Q-state Potts model
on a Bethe lattice are located on arcs of circles with the radius depending on
$Q$ and temperature for $Q>1$.  Complex magnetic field metastability regions are 
studied for the $Q>1$ and $0<Q<1$ cases. The Yang-Lee edge singularity
exponents are calculated for both 1D and Bethe lattice Potts models. The
dynamics of metastability regions for different values of $Q$ is studied
numerically. \\    
PACS number(s): 05.50.+q, 05.10.-a

\end{abstract}

\section{Introduction}

\noindent

The $Q$-state Potts model plays an important role in the general theory of
phase transitions and critical phenomena~\cite{Wu}. It was initially defined
for an integer $Q$ as a generalization of the Ising model ($Q=2$) to
more-than-two components~\cite{Potts}. Later on, it was shown that the Potts
model  for non-integer values of $Q$ may describe the properties of a number of 
physical systems such as dilute spin glasses~\cite{Aharony}, gelatation
and vulcanization of branched polymers ($0<Q<1$)~\cite{Lubensky}. Also it was
shown that the bond and site percolation problems could be  formulated in
terms of Potts models with pair and multisite interactions in the $Q=1$ limit.

In 1952, for the first time, Lee and Yang \cite{Yang} in their famous papers 
studied the distribution of zeros of the partition function considered as a
 function of a complex magnetic filed ($e^{-\frac{2H}{kT}}$ activity, $H$ is a 
magnetic field). They proved the circle theorem which states that zeros of 
the partition function of an Ising ferromagnet lie on unit circle in the
complex activity plane ({\it Yang-Lee zeros}). After these pioneer works
of Lee and Yang, Fisher~\cite{Fisher}, in 1964, initiated the study of
partition function zeros in the complex temperature plane ({\it Fisher
zeros}). These methods were then extended to other types of interactions
and found wide applications~\cite{Arndt}. 

Recently, much attention was drawn to the problem of Yang-Lee and
Fisher zeros  of the $Q$-state Potts model for both integer and non-integer
values of $Q$. The microcanonical transfer matrix method was used to
study the Yang-Lee and Fisher zeros of the non-integer $Q$-state Potts model
in two and three dimensions~\cite{Kim}-\cite{Matveev}.

Derrida, De Seze and Itzykson~\cite{Derrida} showed for the first time that 
the Fisher zeros in hierarchical lattice models are just the Julia set 
corresponding to the renormalization transformation. They found a fractal 
structure for the Fisher zeros in $Q$-state Potts model on the diamond lattice. 
Recently, Monroe investigated Julia sets of the Potts model on recursive lattices%
~\cite{Monroe2}.
He found that the box counting fractal dimension of the Julia set of the
governing recurrence relation is a minimum at a phase transition. This gives an 
alternative criterion for studying phase transitions in models defined on recursive 
lattices.   

In 1994, Glumac and Uzelac~\cite{Glumac}, using the transfer matrix method, 
analytically studied the distribution of Yang-Lee zeros for the one-dimensional
ferromagnetic Potts model with arbitrary and continuous  $Q\geq 0$. For $0<Q<1$ they
obtained that for high temperatures the Yang-Lee zeros lie on a real interval and
for low temperatures these are located partially on the real axis and
in complex conjugate pairs on the activity plane. Later on, Monroe
investigated this model by means of the dynamical systems
approach and confirmed that for $Q<1$ there is a real interval of Yang-Lee
zeros~\cite{Monroe}. Then, Kim and Creswick found that
for $Q>1$ the Yang-Lee zeros lie on a circle with radius $R$, where $R<1$ for
$1<Q<2$, $R>1$ for $Q>2$ and $R=2$ for $Q=2$~\cite{Creswick}. However, it is 
not clear yet what is the location of Yang-Lee zeros for
$0<Q<1$ at low temperatures~\cite{Glumac}. 

In this paper the Yang-Lee zeros of the one-dimensional and Bethe lattice Q-state
Potts models are studied using the dynamical systems approach. It is shown that
for the one-dimensional Potts model the partition function becomes zero when the 
corresponding recurrence relation has neutral fixed points for a given value 
of magnetic field.
Using this correspondence between zeros of the partition function and neutral 
fixed points of the recurrence relation the Yang-Lee zeros of both ferromagnetic and 
antiferromagnetic Potts models are completely studied analytically. The location of 
Yang-Lee zeros of the ferromagnetic Potts model for $0<Q<1$ is found.
Also, formulas for the density of Yang-Lee zeros are derived and edge singularity exponents
are calculated. For the Potts model on a Bethe lattice it is shown that the Yang-Lee 
zeros are located on a phase coexistence line in the complex magnetic field plane.
Here, the phase coexistence line is defined as a line in the complex magnetic field plane,
where the absolute values of derivatives of the recurrence relation in two 
attracting fixed points are equal. It is noteworthy that, Monroe~\cite{Monroe3} used a 
similar criterion for studying critical properties of the Potts model on recursive lattices. 
For the Bethe lattice case an analytical study of Yang-Lee zeros is performed also. A formula for 
Yang-Lee  edge singularity points is derived and edge singularity exponents are calculated. 
Our analitical treatment confirmed the results of numerical calculations. Further, metastability
regions in a complex magnetic field plane are investigated. It is shown that the border
of a metastability region may be found from the condition of existence of a neutral fixed
point of the governing recurrence relation.

The structure of this paper is as follows,
in Section 2 an exact recurrence relation (Potts-Bethe mapping) for the
$Q$-state Potts model on a Bethe lattice is derived. Applying the theory of dynamical
systems to the problem of phase transitions it is shown that critical points 
may be associated with neutral periodical points of the corresponding
mapping. In section 3 the Yang-Lee zeros and edge singularities of
ferromagnetic and antiferromagnetic Potts models are studied analytically for
non-integer $Q$, considering a 1D lattice as a Bethe lattice with coordination number
$\gamma=2$. In Section 4 the Yang-Lee zeros and edge singularities of the
Potts model on a Bethe lattice with coordination number $\gamma>2$ are studied
numerically. In the last section the dynamics of complex magnetic
metastability regions is studied numerically and the explanation of results is given. 

\section{The $Q$-state Potts Model on the Bethe lattice}

The $Q$ state Potts model in the magnetic field is defined by the Hamiltonian 
\begin{equation}
\mathcal{-\beta H}=J\sum_{<i,j>}\delta (\sigma _{i},\sigma
_{j})+h\sum_{i}\delta (\sigma _{i},0), 
\label{ham}
\end{equation}
where ${\sigma }_i$ takes the values $0$, $1$, $2$, $\ldots$, $Q-1$ and $\beta=1/kT$. 
The first sum in the r.h.s. of (\ref{ham}) goes over all edges and the second one 
over all sites on the lattice. 
\begin{figure}
\center
\epsfbox{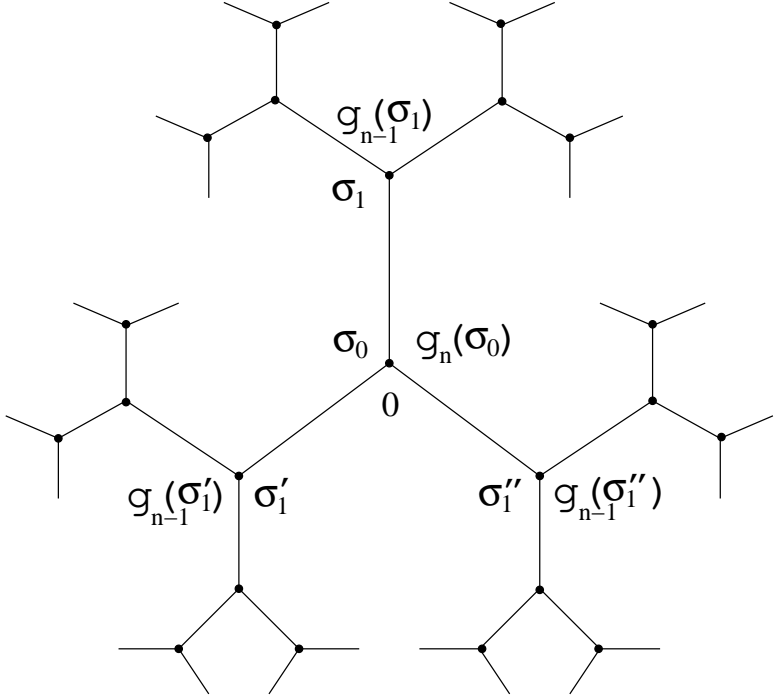}
\caption{\small The Bethe lattice with coordination number $\gamma=3$. 
 \label{Bethe:f}}
\end{figure}
The partition function of the model is given by
\[
\mathcal{Z}=\sum_{\{\sigma \}}e^{-\mathcal{\beta H}}, \label{part1}
\]
where the summation goes over all configurations of the system.

Using the recursive structure of the Bethe lattice (Fig.~\ref{Bethe:f}) one
can derive an exact recurrence relation and apply  the theory of dynamical systems
to investigate the thermodynamical properties of models defined on 
it~\cite{Ghulg1}. Cutting the lattice at the central site $0$ one will
obtain $\gamma$ interacting $n$th generation branches~\cite{Baxter}. By
denoting the partition function of the $n$th generation branch with the basic
site $0$ in the state $\sigma_0$ as $g_n(\sigma_0)$ the partition function may be 
written as follows
\begin{equation}
\mathcal{Z}_n=\sum_{\{\sigma _0\}}exp\left\{ h\delta (\sigma _0,0)\right\} {%
[g_n({\sigma _0})]}^\gamma ,  \label{part2}
\end{equation}
where ${\sigma }_0$ is the Potts variable at the central site $0$ of the
lattice (Fig.~\ref{Bethe:f}). Applying the "cutting" procedure to a $n$th generation branch
one can derive the recurrence relation for $g_n(\sigma_0)$
\begin{equation}
g_n({\sigma }_0)=\sum_{\{\sigma _1\}}exp\left\{ J\delta (\sigma _0,\sigma
_1)+h\delta (\sigma _1,0)\right\} {[g_{n-1}({\sigma }_1)]}^{\gamma -1}.
\label{grecur}
\end{equation}
Introducing the notation 
\begin{equation}
x_n=\frac{g_n(\sigma \neq 0)}{g_n(\sigma =0)},  \label{xn}
\end{equation}
one can obtain the Potts-Bethe map from the Eq(\ref{grecur})
\begin{equation}
x_n=f(x_{n-1}),\qquad f(x)=\frac{\mu +(z+Q-2)\,x^{\gamma-1}}%
{z\,\mu+(Q-1)\,x^{\gamma-1}}. 
\label{xrecgen}
\end{equation}
where $\mu =e^h$, and $z=e^J$. 

The magnetization of the central site for a Bethe lattice of $n$ 
generations may be written as
\begin{equation}
M_n=%\langle \delta(\sigma_0,0)\rangle 
{\mathcal Z}_n^{-1}\sum_{\{\sigma_0\}}\delta(\sigma_0,0)\,e^{\mathcal{-\beta H}}%
= \frac{\mu}{\mu +(Q-1)\,x_n^\gamma}. 
\label{magnet} 
\end{equation}
Instead of $M(\mu)$ in Eq.(\ref{magnet})
it is more convenient to consider the function $\bar{M}(\mu)=2M(\mu)-1$, 
which has the same analytical properties as $M(\mu)$ and give correct magnetization
for the Ising model $(Q=2)$
\begin{equation}
\bar{M}(\mu)=\frac{\mu-(Q-1)x^\gamma}{\mu+(Q-1)x^\gamma}.
\label{magnet:bar}
\end{equation}
In the following the formulas in Eqs.(\ref{xrecgen}) and
(\ref{magnet}) will be generalized to non-integer values of $Q\geq 0$.

Let us now give some definitions and briefly discuss the problem of phase
transitions on recursive lattices in terms of dynamical systems theory. 
The point $x^*$ is called a periodical point with period $n$ of the mapping 
$x_n=f(x_{n-1})$ if it is a solution to the equation $f^k(x)=x$. Here
$f^n$ means a superposition $f^n\equiv f\circ f\circ \cdots \circ f$. If $k=1$,
$x^*$ is called a fixed point. To analyze the stability of a periodic
point $x^*$ of period $k$, i.e. either iterations of $f(x)$ tend to the
periodical point or move away from it, one should compute the derivative of
$f^k$, $\lambda = (f^k)^\prime(x)$ $\left({}^\prime=\frac{d}{dx}\right)$, at
this point. A periodic point $x^*$ is: (1) attracting (stable) if
$\left|\lambda\right|<1$; (2) repelling (unstable) if $\left|\lambda\right|>1$
and (3) neutral (indifferent) if  $\left|\lambda\right|=1$. 

The thermodynamic properties of models defined on recursive lattices may be
investigated by studying the dynamics of the corresponding recursive function. 
For the ferromagnetic Ising model, for example, this may be done as follows. 
For high temperatures ($T>T_c$) the recursive function Eq.(\ref{xrecgen}) has only
one attracting fixed point corresponding to a stable paramagnetic state. For low
temperatures ($T<T_c$) the recursive function Eq.(\ref{xrecgen}) has two attracting
fixed points. In the absence of a magnetic field these two attracting fixed points
correspond to two possible ferromagnetic states with opposite magnetizations. It is
well known that for $h=0$ and $T<T_c$ the system undergoes a first order phase
transition and the condition $\left|\lambda_1\right|=\left|\lambda_2\right|$ is satisfied,
where $\left|\lambda_{1,2}\right|$ are derivatives of $f(x)$ in these attracting
fixed points. In the presence of a magnetic field the stable state corresponds to the
fixed point with maximum $\left|\lambda\right|$ and the magnetization in this state
has the same direction as the external magnetic field. The other fixed point
corresponds to a metastable state, which may be achieved by a sudden reversal in the
sign of the magnetic field. The boundary of the metastability region may be found
from the condition that one of the fixed points becomes neutral. For more details
about the dynamics of metastable states see the book by Chaikin and Lubensky~\cite%
{Chaikin}. The critical temperature corresponds to the values of the magnetic field and
temperature when the fixed points of recursive function $f(x)$ are neutral and repelling.
The values of external parameters (temperature,
magnetic field, etc.) at which the recursive function $f(x)$ has a neutral
periodical point of period $k$ may be obtained from the following system of
equations
\begin{equation}
\left\{
\begin{array}{l}
f^k(x)=x \\ 
\left| {f^k}\,^\prime(x)\right|=1.

\label{neutral}
\end{array}
\right.
\end{equation}
Thus, for the models defined on recursive lattices, critical points and the boundary
of metastability regions may be associated with neutral periodical points of the
model's mapping. In the following section the Yang-Lee zeros of
the 1D $Q$-state Potts model are investigated in terms of neutral fixed points. 

\section{The Yang-Lee zeros of 1D $Q$-state Potts model}
\noindent

A one-dimensional lattice may be considered as a particular case of the Bethe
lattice with coordination number $\gamma=2$. In this case the Bethe-Potts
mapping (\ref{xrecgen}) becomes a M\"{o}bius transformation, i.e. a rational map
of the form
\[
R(x)=\frac{a\,x+b}{c\,x+d},\qquad ad-bc\neq0,
\]
where
\[
R(\infty)=a/c, \qquad R(-d/c)=\infty,
\]
if $c\neq 0$, while $R(\infty)=\infty$ when $c=0$.
The dynamics of
such maps is rather simple~\cite{Beardon}: If $f(x)$ is a M\"{o}bius transformation
with two fixed points, then either $f^n(x)$ converge to one of the
fixed points of $f(x)$ (one of the fixed points is attracting and the other is 
repelling), or they move cyclically through a finite set of points, or they form a
dense subset of some circle (both fixed points are neutral). It follows from the dynamics of
the M\"{o}bius transformation that phase transitions correspond to values of the 
temperature and magnetic field where the recursive function $f(x)$ has only neutral fixed
points. Thus, to study the Yang-Lee zeros one should study the system (\ref{neutral})
to find the values of $\mu$ at which this system has solutions for $k=1$. 
 For neutral fixed points the system (\ref{neutral}) may be written in the form 
\begin{equation}
\left\{
\begin{array}{l}
f(x)=x \\ f^\prime(x)=e^{i\phi},\quad \phi\in [0,2\pi],
\label{1dneutral}
\end{array}
\right.
\end{equation}
Excluding $x$ from the equations of the system (\ref{1dneutral}) after
some algebra one can find the equation of phase transitions in the form
\begin{equation}
z^2\mu^2-2\left[(z-1)\,(z+Q-1)\,\cos\phi+1-Q\right]\mu+\left(z+Q-2\right)^2=0.
\label{phasetr}
\end{equation}
where $\phi\in [0,2\pi]$. Since $cos\phi$ is an even function of $\phi$, 
we may restrict ourselves to $\phi\in[0,\pi]$. Thus, for given $z$
and $Q$ the equation (\ref{phasetr}) is a parametric equation of Yang-Lee zeros,
where $\phi$ is a parameter. Analyzing Eq.(\ref{phasetr}) one can find the
location of Yang-Lee zeros of the Potts model. It is noteworthy that using
this equation one can study the Fisher and Potts zeros%
\footnote{Zeros of the partition function considered as a function of complex $Q$.}
in the same way\footnote{The results will be published elsewhere.}.

First of all note that Eq.(\ref{phasetr}) is a quadratic equation of $\mu$ with real
coefficients. Hence, solutions to this equation lie either on the real axes or in
complex conjugate pairs on a circle with radius 
$R=\left|\mu\right|=\frac{\left|z+Q-2\right|}{z}$ for any
$\phi\in[0,\pi]$ (See also~\cite{Creswick}). The solutions to the
Eq(\ref{phasetr}) can be written in the form
\begin{equation}
\mu_{1,2}=A\,\Big[2\,\cos^2{\textstyle\frac{\phi}{2}}-B\pm2\,\sqrt{\cos^2%
{\textstyle\frac{\phi}{2}}%
\left(\cos^2{\textstyle\frac{\phi}{2}}-B\right)}\Big], \label{sol}
\end{equation}
where
\begin{equation}
A=\frac{(z-1)\,(z+Q-1)}{z^2} \quad\mbox{and}\quad
B=\frac{z(z+Q-2)}{(z-1)\,(z+Q-1)}.
\end{equation}

Let us now study the Yang-Lee zeros of ferromagnetic Potts model, i.e.
$J>0$ and $z>1$. For $Q>1$ one can easily find that $B>1$, hence, all
solutions (\ref{sol}) are complex conjugate and lie on an arc of circle with
radius $R=\frac{z+Q-2}{z}$. If $1<Q<2$ then $R<1$ (the arc lies inside the unit
circle), if $Q>2$ then $R>1$ (the arc lies outside the unit circle~\cite{Creswick})
and $R=1$~\cite{Yang} for $Q=2$ (Ising model). Writing $\mu$
in the exponential form $\mu=R\,e^{i\,\theta}$ one can find 
\begin{equation}
\cos\frac{\theta}{2}=\sqrt\frac{(z-1)\,(z+Q-1)}{z\,(z+Q-2)}\,\cos\frac{\phi}{2}.
\label{gap}
\end{equation}
From Eq.(\ref{gap}) one can see that there are no solutions with arguments
in the interval $0<\theta<\theta_0$, where $\theta_0=2\,\arccos \sqrt{B^{-1}}$. 
This is the well known gap~\cite{Kortman} in the distribution of Yang-Lee zeros. 
It points to the absence of phase transitions in an 1D ferromagnetic Potts model for
$Q>1$ at any real temperature. This is in good agreement with recent studies
by Glumac and Uzelac~\cite{Glumac}, and Kim and Creswick~\cite{Creswick}, where
the Yang-Lee zeros of the 1D Potts model was studied by using the transfer matrix
method. Comparing our formula (\ref{gap}) with the corresponding formula
(14) of Ref.~\cite{Creswick} one can see that the argument $\phi$ of the
derivative in our method is nothing but the difference in the arguments of
two maximal eigenvalues in the transfer matrix method. 
It follows from the Eqs.(\ref{sol}) and (\ref{gap}) that the Yang-Lee edge fields
correspond to $\phi=0$ and have the form
\begin{equation}
\mu_\pm=\frac{1}{z^2}\left\{\sqrt{(z-1)(z+Q-1)}\pm\sqrt{1-Q}\right\}^2,
\label{edge}
\end{equation}
and $\mu_\pm$ are complex for $Q>1$.

$B<1$ when $Q<1$ and it is positive or negative depending on $z$.  
Hence,  $B$ is negative when $z\leq2-Q$ ($Q<1$), and all values of $\mu$'s in
(\ref{sol}) are real and lie between the $\mu_-$ and $\mu_+$ where
$\mu_\pm>0$.  

$0<B<1$ when $z>2-Q$ ($Q<1$), and all values of $\mu$'s in (\ref{sol}) are either real or
complex depending on $\phi$. For $0<\phi<\phi_0$, where
$\phi_0=2\,\arccos\sqrt{B}$, solutions (\ref{sol}) are real and lie in the
interval $[\mu_-,\mu_+]$. For $\phi_0<\phi<\pi$ the solutions
(\ref{sol}) are complex conjugate and lie on the circle with radius $R=\frac{z+Q-2}{z}$.

For $Q>1$, the differentiation of both sides of Eq.(\ref{gap}) with 
respect to $\mu$ and $\phi$ gives the density of Yang-Lee zeros $g(\theta)$ in the form
\begin{equation}
g(\theta)=\frac{1}{2\,\pi}\frac{|\sin\frac{\theta}{2}|}{\sqrt{\sin^2\frac{\theta}{2}
 - \sin^2\frac{\theta_0}{2}}}.
\label{density1}
\end{equation}
From Eq.(\ref{density1}) it follows that the density function $g(\theta)$ 
for $Q>1$ diverges in the gap points $\mu_\pm$ with the exponent $\sigma=-\frac 1%
2$, i.e. $g(\theta)\propto |\theta-\theta_0|^{\sigma}$ when $\phi\to 0$ or 
$\theta\to\theta_0$.

For $Q<1$, the corresponding
density function $g(\mu)$ may be obtained by differentiation of both sides of 
Eq.(\ref{phasetr})
\begin{equation}
g(\mu)=\frac{1}{2\,\pi\mu}\frac{|\mu-\sqrt{\mu_+\mu_-}|}%
{\sqrt{(\mu_+-\mu)(\mu-\mu_-)}}.
\label{density2}
\end{equation}  
One can see that $g(\mu)$ diverges in the points $\mu_\pm$, i.e. 
$g(\mu)\propto |\mu-\mu_\pm|^{\sigma}$, with the exponent 
$\sigma=-\frac12$. Thus, for $Q<1$ the density function of Yang-Lee zeros of the  1D
$Q$-state Potts ferromagnetic model has singularities only at the points $\mu_\pm$ 
[Eq.(\ref{edge})] with the edge singularity exponent $\sigma=-\frac12$. The same is
true for the antiferromagnetic case.

\begin{figure}[h]
%\hspace{-1.5cm}
\epsfxsize=16cm
\epsfbox{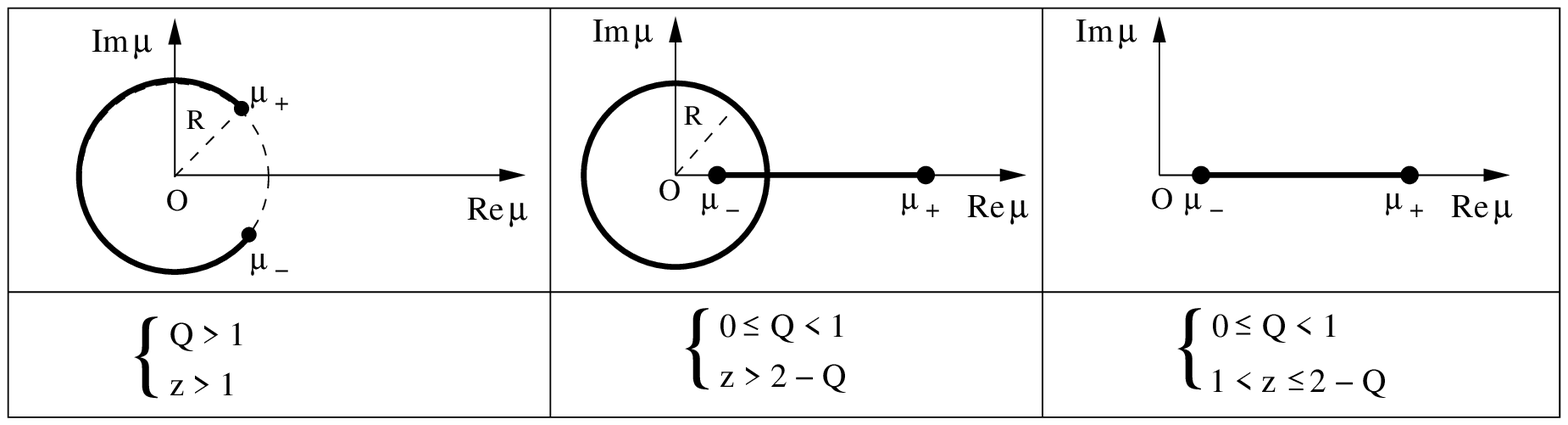}
\caption{A schematic representation of the Yang-Lee zeros of 1D ferromagnetic
Potts model. Here $R=\frac{z+Q-2}{z}$ and $\mu_\pm$ are defined in (\ref{edge}).
\label{1d:ferro}}
\end{figure}

Glumac and Uzelac~\cite{Glumac} considered for $Q<1$ case the contribution of the third
eigenvalue of the transfer matrix ($\lambda_2$ in their notations). We want to note that for the
1D Potts model the transfer matrix method gives three and more eigenvalues
 only for $Q>2$, hence, the third and other eigenvalues should be neglected for
$Q<2$. In this case the study of two maximal eigenvalues gives the same
results as our method. The summary of results of the 1D ferromagnetic Potts
model is given in Fig.~\ref{1d:ferro}. 
It is interesting to note that the argument $\phi$ of the derivative in 
Eq.(\ref{1dneutral}) corresponds to the argument of the maximal eigenvalue in 
the transfer matrix method.

\begin{figure}[h]
%\hspace{-1.5cm}
\epsfxsize=16cm
\epsfbox{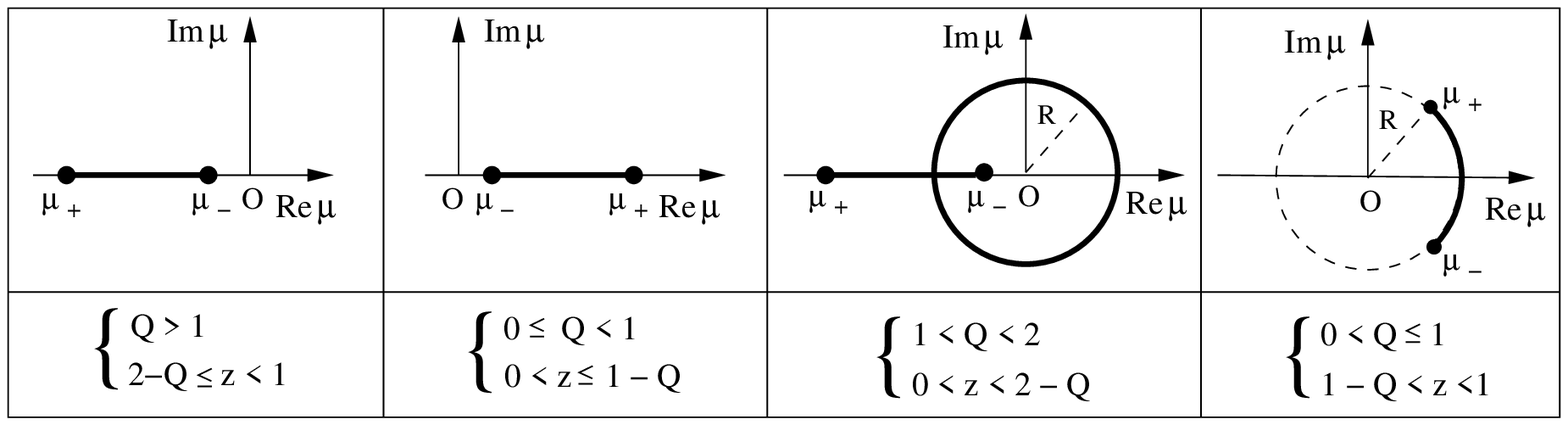}
\caption{A schematic representation of the Yang-Lee zeros of 1D
antiferromagnetic Potts model. Here $R=\frac{2-Q-z}{z}$ and $\mu_\pm$ are 
defined in (\ref{edge}). \label{1d:antiferro}}
\end{figure}

The antiferromagnetic case may be studied in
the same manner. The results are shown in Fig.~\ref{1d:antiferro}.

\section{Yang-Lee zeros of the $Q-$ state Potts model on a Bethe lattice ($Q>1$)}
\noindent

\begin{figure}[h]
\epsfbox{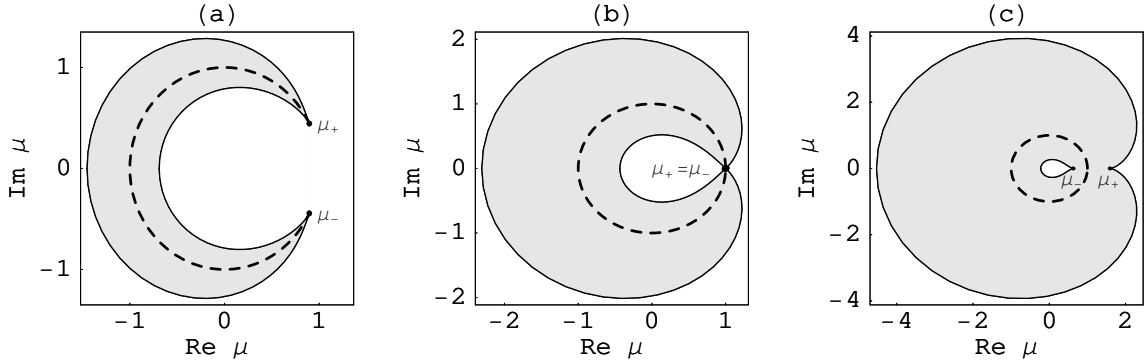}
\caption{Metastability regions and Yang-Lee zeros of the $Q$-state Potts model
on the Bethe lattice with coordination number $\gamma=3$ and $Q=2$. The solid
lines correspond to the boundary of metastability regions (gray filled areas).
Dashed lines present an arc or circles of radius 
$R_\mu^2=\frac{(z+Q-2)^\gamma}{z^\gamma(Q-1)^{\gamma-2}}$ and correspond 
to Yang-Lee zeros. 
(a) $T>T_c$ ($z=1.8$), $\mu_\pm$ are Yang-Lee edge singularity points; 
(b) $T=T_c$ critical point ($z=z_c=3$); 
(c) $T<T_c$ ($z=6$), $\mu_\pm$ are spinodal points of the model. 
For more details see the text.
\label{metast:border}}
\end{figure}

Let us, at first, consider the ferromagnetic Potts model on the Bethe lattice 
with coordination number $\gamma=3$. In this case the system (\ref{neutral}) may be
studied analytically for neutral fixed points ($k=1$). The exclusion of $x$ from both
equations (\ref{neutral}) gives the following equation
\begin{equation}
z^3(Q-1)\mu^2-[2(Q-1)y(4\cos\phi+1)+y^2(4e^{i\phi}\sin^2{\textstyle\frac{\phi}{2}}+%
1)-2(Q-1)^2]\mu+(z+Q-2)^3=0.
\label{metastable}
\end{equation}
where $y=(z-1)(z+Q-1)/2$ and $\phi\in[0,2\pi]$. This equation describes the border
of the metastability region in the complex $\mu$ plane (Fig.~\ref{metast:border}).
The dashed areas in Fig.~\ref{metast:border} show the metastability regions for
different temperatures. Inside the metastability region there are two attracting
fixed points and there is only one outside of it. The other fixed
points are repelling. Note that such a behavior is valid for any $\gamma$. 
At $\mu_\pm$ at least two of the fixed points become neutral.
It will be shown below that at $\mu_\pm$ the magnetization function
(\ref{magnet:bar}) is singular and  $\mu_\pm$ correspond to the Yang-Lee edge
singularity points. These are solutions to the equation (\ref{metastable}) for
$\phi=0$ and the critical temperature may be obtained from the condition
$\mu_+=\mu_-$. It follows from Eq(\ref{metastable}) that the edge singularity points
lie on a circle with radius $R_\mu^2=\frac{(z+Q-2)^3}{z^3(Q-1)}$. Since there is 
no phase transition on the boundary of a metastability region it will not give rise to
zeros of the partition function~\cite{Chaikin}. The metastability region in the
complex $\mu$-plane points to the existence of the first order phase transition for
complex magnetic fields. The problem of finding the Yang-Lee zeros of models with
first order phase transitions attracted much attention for many years. Recently,
Biskup et al., showed that the position of partition function zeros is related
to the phase coexistence lines in the complex planes~\cite{Biskup}. 
Dolan et al., used this approach to study Fisher zeros for Ising and Potts
models on non-planar ('thin') regular random graphs. It is interesting to note
that the locus of Fisher zeros on a Bethe lattice is identical to the corresponding
random graph~\cite{Dolan}. For our models the phase coexistence line is defined as a
line in the complex plane, where the absolute values of the recursive  
function  derivatives in two attracting fixed points are equal(see also~\cite{Monroe3}). 
Our numerical study showed that the phase coexistence line for $T>T_c$ is an
arc of a circle with radius $R_\mu$ ending at the edge singularity points (the dashed
line in Fig.~\ref{metast:border}). 

Let us now study the analytical properties of the magnetization function~
(\ref{magnet:bar}) to prove that the edge singularity points $\mu_\pm$ correspond to 
its singularities. 
The fixed point equation of the Potts-Bethe mapping (\ref{xrecgen}) may be written
in the form
\begin{equation}
\mu=x^{\gamma-1}\frac{(Q-1)\,x-(z+Q-2)}{1-z\,x}.
\label{mux}
\end{equation}
For continuity at $\mu\neq 0$, $x$ is defined to be equal to $(Q-1)/(z+Q-2)$ at
$\mu=0$. Considering $x$ as function of $\mu$, one can study the singularities
of $x(\mu)$ that also correspond to singularities of $\bar{M}(\mu)$. The singular
points of $x(\mu)$ are $\mu=0$, $ \mu=\infty$ and $\mu(x_\pm)$, $x_\pm$ being the
points, where the derivative $\partial x/\partial\mu$ is infinite. $x_\pm$ satisfies the
equation 
\begin{equation}
x^2+\frac{z(z+Q-2)(2-\gamma)-\gamma(Q-1)}{z(Q-1)(\gamma-1)}x+\frac{z+Q-2}{z(Q-1)}=0
\label{edgex}
\end{equation}
Eq.(\ref{edgex}) is obtained after differentiation both sides of Eq.(\ref{mux}) 
with respect to $\mu$ and $x$ from the condition that $\partial\mu/\partial x$ 
vanishes. Note that Eq.(\ref{edgex}) may be derived from the system  
(\ref{neutral}) for $k=1$ and $\phi=0$ by excluding $\mu$ from both 
of the equations. At the singularity points of $\bar{M}(\mu)$ the recursive
function has two neutral fixed points with $\phi=0$ and the others are repelling.

The critical temperature may be obtained from the condition $\mu_+=\mu_-$ or 
setting the discriminant of the quadratic equation (\ref{edgex}) to zero
\begin{equation}
z_c=\left\{ 
\begin{array}{cl}
\frac12\left(2-Q+\sqrt{(Q-2)^2+4(Q-1)\gamma^2/(\gamma-2)^2})\right) %
& \mbox{for $Q>1$}, \\ %
1-Q & \mbox{for $Q<1$}. 
\end{array}
\right.
\end{equation}
For $Q>1$ and $1<z<z_c$ the solutions $x_\pm$ are complex conjugate numbers with
modulus $R_x=\sqrt\frac{z+Q-2}{z(Q-1)}$ and for $z>z_c$ they become real valued. For $Q<1$
the solutions $x_\pm$ are always real numbers. Substituting $x_\pm=R_x%
e^{i\alpha_\pm}$ into Eq.(\ref{mux}) one finds after some algebra  
\begin{equation}
\mu_\pm=R_\mu e^{i[\tilde{\theta}_\pm+\alpha_\pm(\gamma-1)]},
\end{equation}
where 
\begin{equation}
R_\mu=R_x^\gamma(Q-1),\quad
\tan{\textstyle\frac{{\tilde\theta}_\pm}{2}}=\frac{1+z\,R_x}{1-z\,R_x}\,%
\tan{\textstyle\frac{\alpha_\pm}{2}}
\quad \mbox{and} \quad 
\cos\alpha_\pm=\frac{(\gamma-2)z^2R_x^2+\gamma}{2(\gamma-1)z\,R_x}.
\end{equation}
Note that the fixed point
equation $f(x)=x$ and the magnetization $\bar{M}(\mu)$ are
invariant under the transformation $G:\{\mu\rightarrow%
\frac {R_\mu^2}{\mu},\,\, x\rightarrow \frac {R_x^2}{x} \}$. Moreover, for the
magnetization function one has
\begin{equation}
\bar{M}(\mu)=-\bar{M}\left(\frac{R_\mu^2}{\mu}\right).
\label{magnet:inv}
\end{equation}

The Yang-Lee zeros of the Potts model on the Bethe lattice may be obtained also
analytically by studying the analytical properties of the magnetization
function~(\ref{magnet:bar}). This was done for the first time by Bessis,
Drouffe and Moussa~\cite{Bessis} for the Ising model on Bethe lattice.
Making use of singularities of $x(\mu)$ in analogy with~\cite{Bessis}
a careful analysis of the analytic properties of the function $\bar{M}_\mu$ Eq.
(\ref{magnet:bar}) for $Q>1$ gives the following picture for Yang-Lee zeros:
(a) For $z<z_c$: $\bar{M}(\mu)$ is analytic in the complex plane cut along an arc of
the circle with radius $R_\mu^2=\frac{(z+Q-2)^\gamma}{z^\gamma(Q-1)^{\gamma-2}}$,
that contains the point $-R_\mu$ and is limited by the points $\mu_\pm$, which are complex
conjugate. Due to relation (\ref{magnet:inv}), the discontinuity along the cut is
real and never vanishes except at the Yang-Lee edge singularity points
$\mu_\pm$. This is in a good agreement with  the Yang-Lee theory~\cite{Yang}. The density of
zeros $g(R_\mu,\theta)$ may be calculated from 
\begin{equation}
\lim_{r\rightarrow R_\mu+} \bar{M}(\mu)\left|_{\mu=re^{i\theta}}\right.-
\lim_{r\rightarrow R_\mu-} \bar{M}(\mu)\left|_{\mu=re^{i\theta}}\right.=-4\pi
g(R_\mu,\theta).
\label{density}
\end{equation}
Our numerical study of Eq.(\ref{density}) shows that close to the edge singularity
points $\mu_\pm=e^{\pm i\theta_0}$, $g(R_\mu,\theta)$ has an exponential behavior
$g(R_\mu,\theta)\propto |\theta-\theta_0|^\sigma$ and $\theta_0(T)\propto%
(T-T_c)^\Delta$ where $\sigma=1/2$ for $z<z_c$, $\sigma=1/3$
$(\sigma=\frac1\delta)$ for $z=z_c$ and $\Delta=\frac32$ $(\Delta=\beta\delta)$.
Our results are in a good agreement with those of~\cite{Fisher2},~\cite{Kortman},
where it was shown that the $\sigma$ exponent is 
universal and always equal to $1/2$ for $T>T_c$ and $\sigma=1/\delta$ for
$T=T_c$ in ferromagnetic models on lattices with spatial dimension $d>6$. 

\begin{figure}[h]
\epsfbox{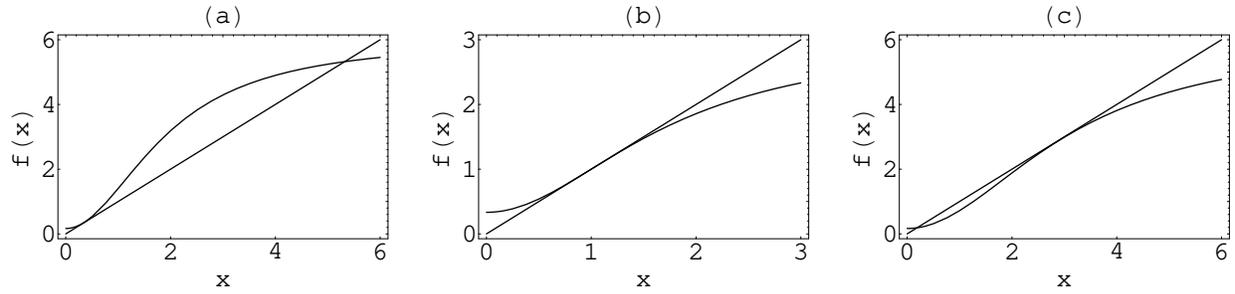}
\caption{Plots of $f(x)$ function (\ref{xrecgen}) indicating the existence of
neutral fixed points  for different temperatures and magnetic fields at $Q=2$ and
$\gamma=3$. (a) $T<T_c$ ($z=6$) and $\exp(h/kT)=\mu_-$; (b) $T=T_c$ ($z=3$) $h=0$;
(c) $T<T_c$ ($z=6$) and $\exp(h/kT)=\mu_+$.\label{fx}}
\end{figure}

(b) For $z>z_c$ the function $\bar{M}(\mu)$ is split into two different functions:
$\bar{M}_+(\mu)$ defined for $|\mu|<R_\mu$ and $\bar{M}_-(\mu)$ defined for 
$|\mu|>R_\mu$. The function $\bar{M}_+(\mu)$ can be analytically continued outside 
the circle $|\mu|=R_\mu$ 
into the $z$ plane cut along the real axes from $\mu(x_+)$ to $\infty$, 
where $x_+$ is the largest root of Eq.(\ref{edgex}). The point $\mu(x_+)$ increases
from unity to infinity when $z$ increases from $z_c$ to infinity. The discontinuity of 
$\bar{M}_+(\mu)$ across the cut is purely imaginary and does not change the sign. Hence,
it has no influence on physical properties of the model and the Yang-Lee zeros
lie on the circle $|\mu|=R_\mu$. The points $\mu_\pm$ correspond to the 
boundary of metastability for real magnetic fields and the plots of the iteration 
function $f(x)$ at these points are given in Fig.~\ref{fx}.

\section{Complex Magnetic Field metastability regions}
\noindent

In the previous section the neutral
fixed points of the recursive function $f(x)$ (\ref{xrecgen}) was considered only.
It was found that the set of values of the magnetic field, for which the recursive
function (\ref{xrecgen}) has at least one neutral fixed point gives the boundary
of the metastability region in the complex $\mu$-plane.
Inside it there are two attracting fixed
points and others are repelling. Numerical experiments show that in the $\mu$-plane
the recursive function Eq.(\ref{xrecgen}) has a complex behavior with
period doubling bifurcations. The question arises: What will happen to the 
metastability region if neutral periodical points of period
$k\geq 1$ are also considered? To answer this question one has to study the system~(\ref{neutral}) for
any $k\geq 1$. Since it is impossible to solve the
system~(\ref{neutral}) directly for $k\gg 1$ and $\gamma\geq 3$ analytically and even
numerically for large $k$ and $\gamma$, the method developed in~\cite{Ghulg2} is
used. It gives a
numerical algorithm for searching neutral periodical points for the recursive
functions like Eq.(\ref{xrecgen}) and is based on the theory of complex dynamical
systems and the well known fact that the convergence of iterations to neutral periodical
points is very weak and irregular, i.e. one has to make a number of iterations
in order to approach a neutral periodic point.
The algorithm is to find all critical points of the recursive
function and investigate the convergence of all the orbits started at critical
points (critical orbits). If all critical orbits converge to any attracting
periodical point one says that the recursive function has only attracting and
repelling periodical points. If at least one of the critical orbits does not converge,
for example, after $n$ iterations, one says that the recursive function has a neutral
periodical point. Of course, the last statement is not rigorous from the strong
mathematical point of view since a weak convergence to an attracting periodical
point is also possible. Depending on the choice of $n$ and $\varepsilon$ (the
accuracy of approaching an attracting periodical point), the resulting picture
on the $\mu$-plane may change. Our simulation experiments show that $n=700$ 
and $\varepsilon=10^{-5}$ are
optimal values and the data generated with this algorithm do not qualitatively
change when $n$ and/or $\varepsilon$ differ from their optimal values. For more
details of the method and the $C++$ program code see~\cite{Ghulg2,Ghulg3}.

One can easily find all critical points of the mapping $f$ from Eq.(\ref
{xrecgen}). The critical points are: $x=0$ with multiplicity $%
\gamma -2$ and $x=\infty $ with multiplicity $\gamma -2$. The degree of our
mapping $f$ is $d=(\gamma -1)$. It may be shown that these are the only critical
points of the mapping $f$ \cite{Ghulg2}. Hence, one has to consider only the
orbits of the points $x_0=0$ and $x_0=\infty$. For numerical calculations
it is convenient to start iterations at the points $x_1=f(0)=1/z$ and 
$x_1=f(\infty)=(z+Q-2)/(Q-1)$.

\begin{figure}[p]
\epsfxsize=15cm
\epsfbox{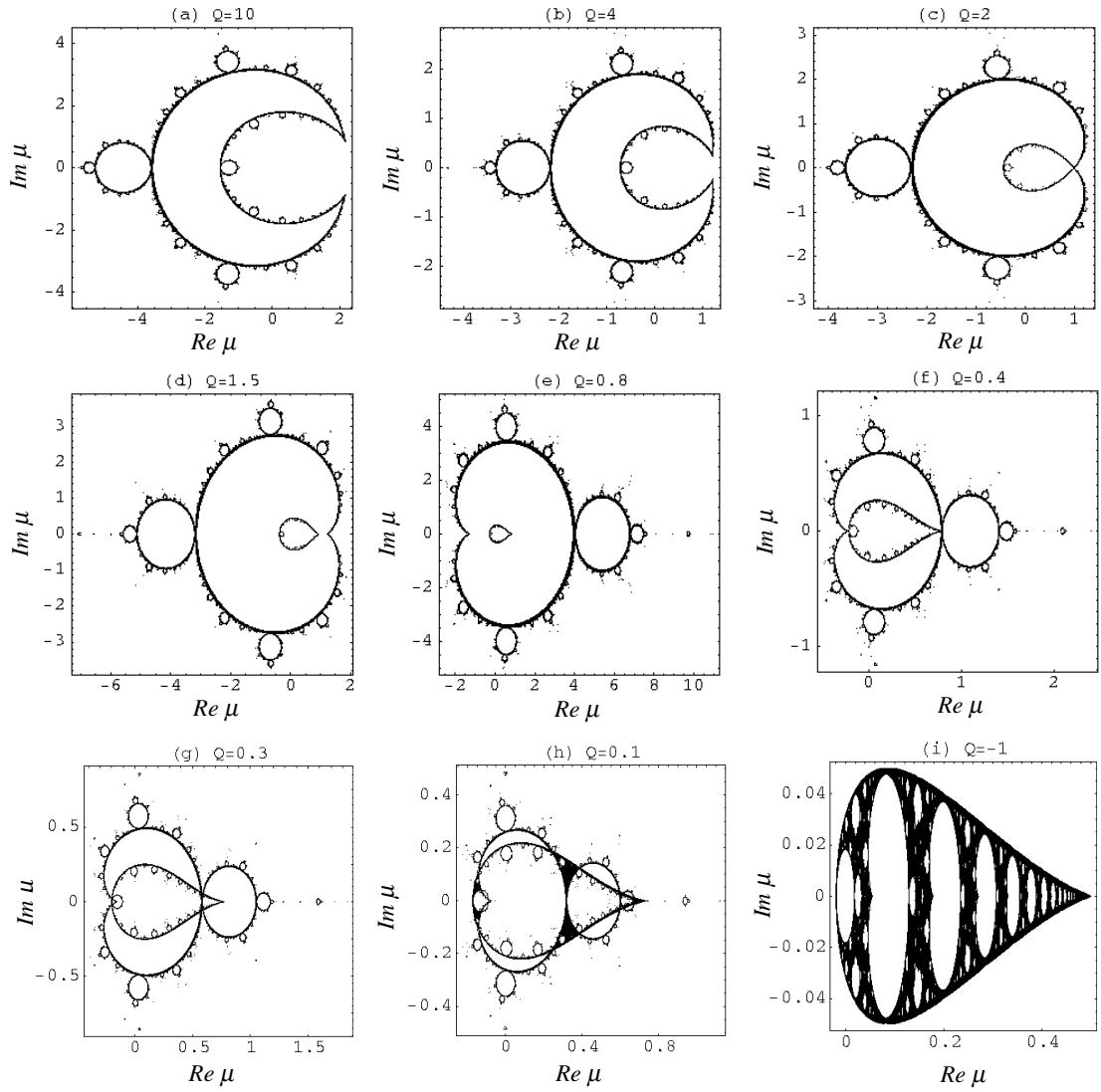}
%\vspace{-10cm}
\caption{The dynamics of metastability regions of the $Q$-state Potts model
on the Bethe lattice with coordination number $\gamma=3$ and $z=3$ for
different values of $Q$. For more details see the text. \label{dyn:metastab}}
\end{figure}

In Fig. \ref{dyn:metastab} the dynamics of metastability regions 
of the $Q$-state Potts model on the Bethe lattice with coordination number
$\gamma=3$ and $z=3$ is shown depending on $Q$. We have experimental evidence that
in white regions all critical orbits converge. Figs. \ref{dyn:metastab}(a)-(d) show
the metastability regions for the case $Q>1$. It is seen that the sets of black points
are similar to the boundary of the Mandelbrot set of the quadratic mapping $z\rightarrow
z^2+c$ (Figs. \ref{dyn:metastab}(c),(d)). This fact is known as the universality of
the Mandelbrot set~\cite{Hubbard}. The metastability region of the Ising model
$(Q=2)$ at the critical temperature is shown in Fig. \ref{dyn:metastab}(c). It 
intersects the positive semi-axis at $\mu=1$, which is an evidence of the existence of
real temperature second order phase transition in conformity 
with exact calculations~\cite{Baxter}. The $Q<1$ case is shown in Figs.
\ref{dyn:metastab}(e)-(i). One can see that Fig. \ref{dyn:metastab}(e) resembles
a mirror reflection of the Mandelbrot set boundary of  (cf. Fig.~\ref{dyn:metastab}(d)-(e)).
It is noteworthy that for $Q=1$ and $\gamma=3$
the Bethe-Potts mapping becomes a quadratic one~\cite{Dolan2}, and our numerical 
method fails~\cite{Ghulg2}.
By lowering $Q$ the metastability regions become more and more complicated
Figs. \ref{dyn:metastab}(f)-(i). Note
that the dynamics of the metastability regions remains the same if one fixes $Q$ and
changes the temperature. 

\section{Conclusions}

We showed that the stability analysis of attracting fixed points of the
recurrence relation, i.e. the condition that the absolute values of derivatives of
the governing recurrence relation in two attracting fixed points are equal at 
a phase transition, may be 
successfully applied to a study of zeros of the partition function. Note that 
in the one-dimensional case the condition of a phase transition is equivalent to the 
condition of existence of neutral fixed points. It will 
be interesting to check the Monroe's conjecture that at a phase transition the box 
counting dimension of the Julia set of the governing recurrence relation is a 
minimum for zeros of the partition function~\cite{Monroe2}. We suppose that the 
box counting dimension of the Julia set of the recurrence relation should be a 
minimum for zeros of the partition function also.This may serve as a new 
criterion for studying zeros of the partition function for models on recursive 
lattices.  

In conclusion we observe that numerical methods proposed in this paper are
generic and may be used for investigation of zeros of the partition function and 
metastability regions of other models on recursive lattices.

\section{Acknowledgments}
\noindent
This work was partly supported by the Grants INTAS-97-347,
ISTC A-102, Dutch NWO: Nato Visitors Grant NB 62-562, and  the Cariplo Foundation.
One of the authors R. G. would like to thank Prof. S. Fauve and Prof. A. Hoekstra for
fruitful discussions and International Centre for Theoretical Physics (ICTP, Italy) for hospitality.

\newpage

\newpage

\end{document}